\documentclass[nature,letterpaper,twocolumn,superscriptaddress,amsmath,amsfonts,amssymb,preprintnumbers,citeautoscript]{revtex4}
\usepackage{dcolumn,graphicx,color,booktabs}
\usepackage{amsmath,bm}
\renewcommand{\figurename}{Fig.}

\begin{document} %\pagestyle{plain}
\title{Direct observation of dispersive lower Hubbard band \\in iron-based superconductor FeSe}

\author{D.\,V.\,Evtushinsky}
\altaffiliation{Current address: BESSY II, Helmholtz-Zentrum Berlin for Materials and Energy, Albert-Einstein-Strasse 15, 12489 Berlin, Germany}
\email{danevt@gmail.com}
\affiliation{Institute for Solid State Research, IFW Dresden, P.\,O.\,Box 270116, D-01171 Dresden, Germany}

\author{M.\,Aichhorn}
\affiliation{Institute of Theoretical and Computational Physics, TU Graz, Petersgasse 16, 8010 Graz, Austria}

\author{Y.\,Sassa}
\affiliation{Laboratory for Solid State Physics, ETH Zurich, CH-8093 Zurich, Switzerland}
\affiliation{Department of Physics and Astronomy, Uppsala University, 751 20 Uppsala, Sweden}

\author{Z.-H. Liu}\author{J. Maletz}
\affiliation{Institute for Solid State Research, IFW Dresden, P.\,O.\,Box 270116, D-01171 Dresden, Germany}

\author{T.\,Wolf}
\affiliation{Karlsruher Institut f\"{u}r Technologie, Institut f\"{u}r Festk\"{o}rperphysik, 76021 Karlsruhe, Germany}
 
\author{A.\,N.\,Yaresko}
\affiliation{Max-Planck-Institute for Solid State Research, Heisenbergstrasse 1, D-70569 Stuttgart, Germany}

\author{S.\,Biermann}
\affiliation{Centre de Physique Theorique, Ecole Polytechnique, CNRS, 91128 Palaiseau Cedex, France}

\author{S.\,V.\,Borisenko}
\affiliation{Institute for Solid State Research, IFW Dresden, P.\,O.\,Box 270116, D-01171 Dresden, Germany}

\author{B.\,B\"{u}chner}
\affiliation{Institute for Solid State Research, IFW Dresden, P.\,O.\,Box 270116, D-01171 Dresden, Germany}
\affiliation{Institut f\"{u}r Festk\"{o}rperphysik, Technische Universit\"{a}t Dresden, D-01171 Dresden, Germany}

\begin{abstract}
\noindent \textbf{Electronic correlations were long suggested not only to be responsible for the complexity of many novel materials, but also to form essential prerequisites for their intriguing properties. Electronic behavior of iron-based superconductors is far from conventional, while the reason for that is not yet understood. Here we present a combined study of the electronic spectrum in the iron-based superconductor FeSe by means of angle-resolved photoemission spectroscopy (ARPES) and dynamical mean field theory (DMFT). Both methods in unison reveal strong deviations of the spectrum from single-electron approximation for the whole 3$d$ band of iron: not only the well separated coherent and incoherent parts of the spectral weight are observed, but also a noticeable dispersion of the lower Hubbard band (LHB) is clearly present.  This way we demonstrate correlations of the most puzzling intermediate coupling strength in iron superconductors.}
\end{abstract}

%Electronic correlations seem to accompany many interesting and useful phenomena in modern materials(, with high temperature superconductivity as one of the most prominent examples). One of the reasons that hinders the understanding of such electronic systems is that they are inbetween well understood limits, non-correlated electrons propagating through the 
%
%weakly and strongly correlated limits, for which analytical solutions are found. 
%
%Hubbard
%
%complex case of intermediate correlations 
%
%
%were long suggested to be  

%\pacs{74.25.Fy, 74.25.Jb, 79.60.-i, 71.20.-b}

\maketitle

\begin{figure}[]
\includegraphics[width=0.6\columnwidth]{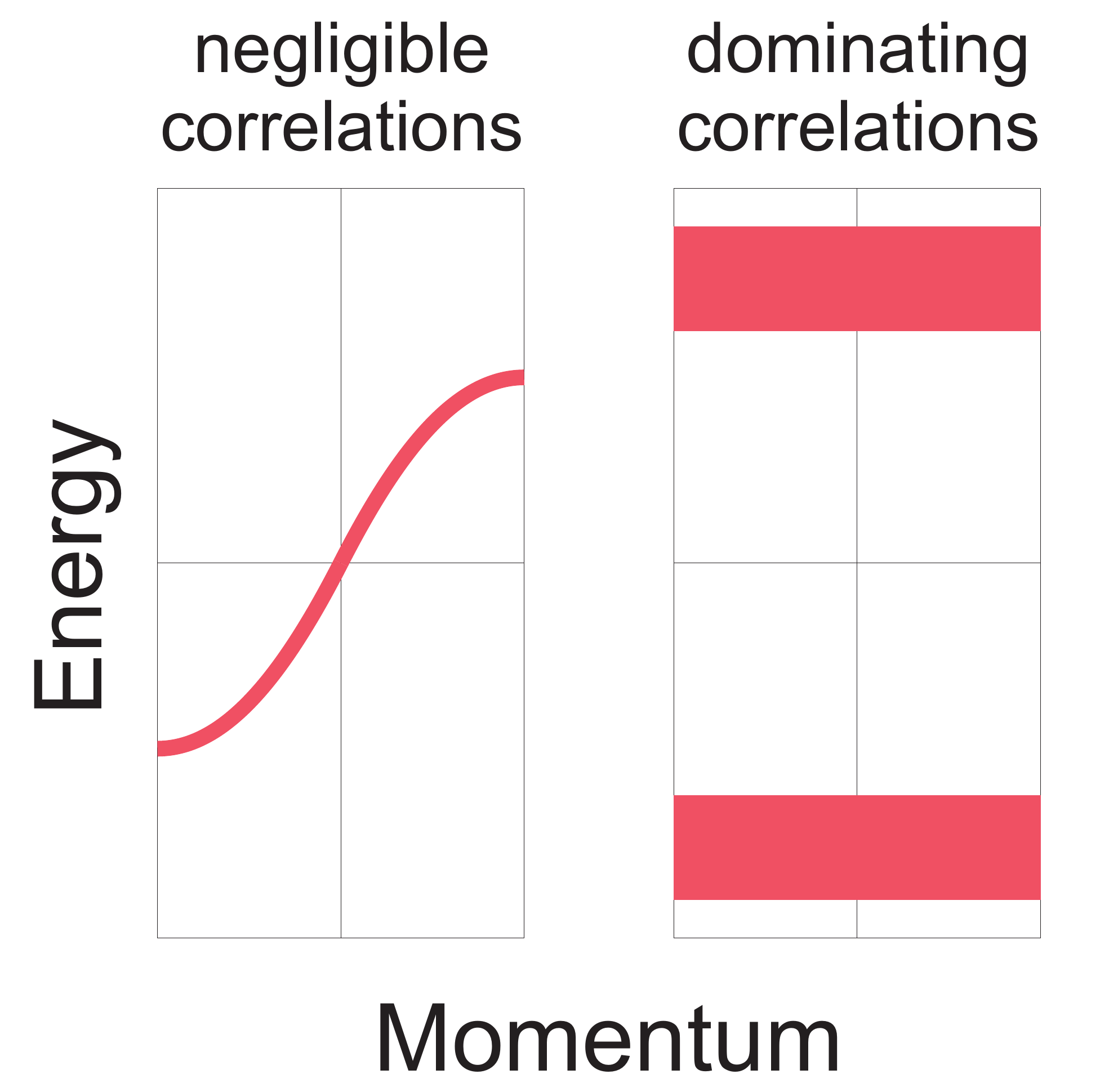}
\vspace{-0.0cm}
\caption{Schematic representation of the electronic bands without and with correlations. Within mean field approach the electronic spectrum of a crystalline solid consists of bands with well-defined dispersion (left), while in the case of strong electron-electron correlations, Hubbard bands are formed (right).}
\vspace{-0.0cm}
\label{one}
\end{figure}

\begin{figure*}[]
\includegraphics[width=0.8\textwidth]{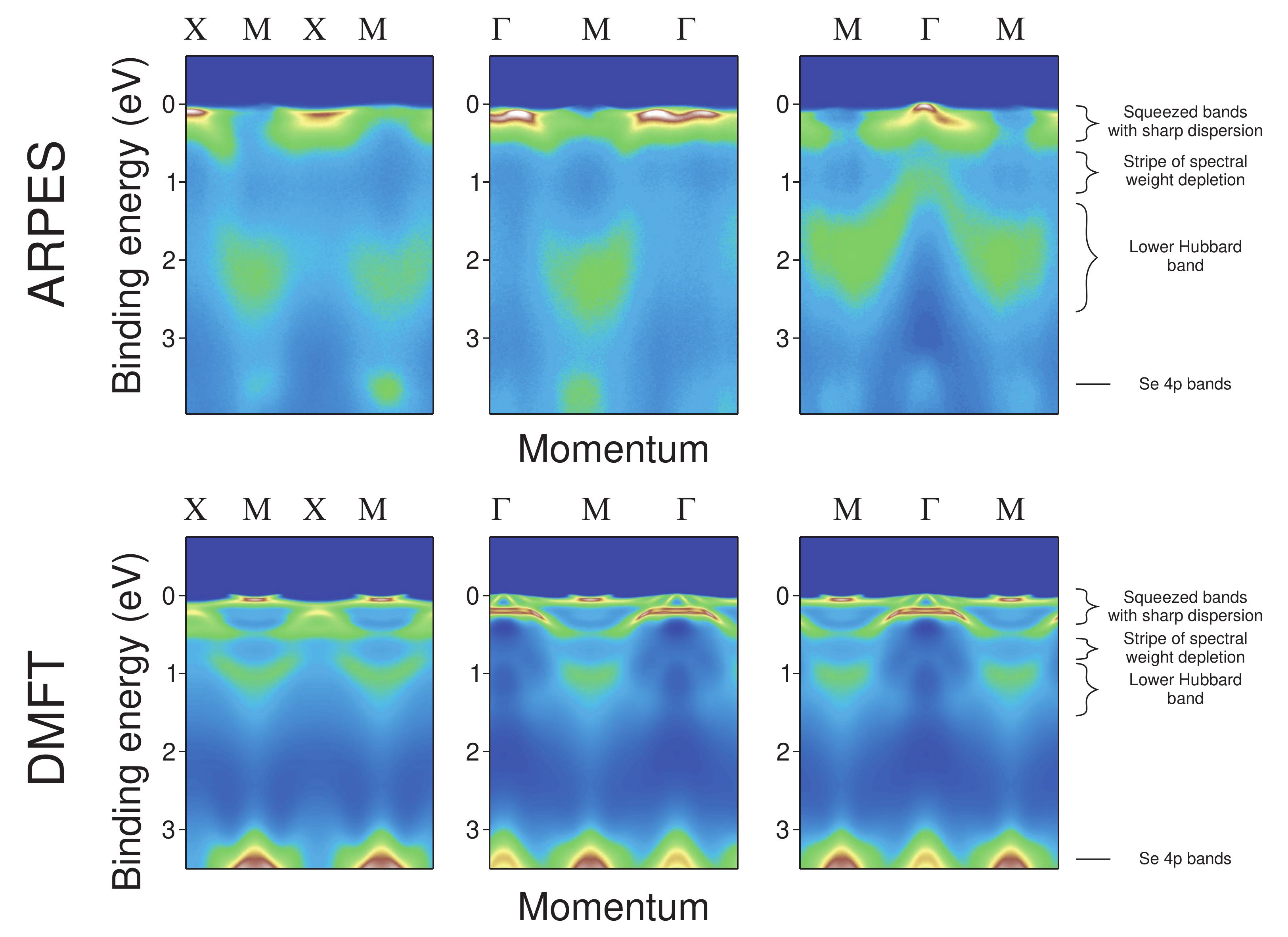}
\vspace{-0.0cm}
\caption{Electronic spectra of FeSe measured by ARPES and calculated by DMFT. Top row: distribution of the spectral weight for 3$d$ band of iron, measured by ARPES along high-symmetry directions in the Brillouin zone. Bottom row: spectral function resulting from DMFT calculations for the same directions. Comparison of DMFT and ARPES shows very good agreement in terms of general structure of the spectrum: In both cases there are squeezed bands with sharp dispersion at the Fermi level, which are separated from the lower Hubbard band by a stripe of spectral weight depletion.}
\vspace{-0.0cm}
\label{two}
\end{figure*}

\begin{figure*}[]
\includegraphics[width=0.75\textwidth]{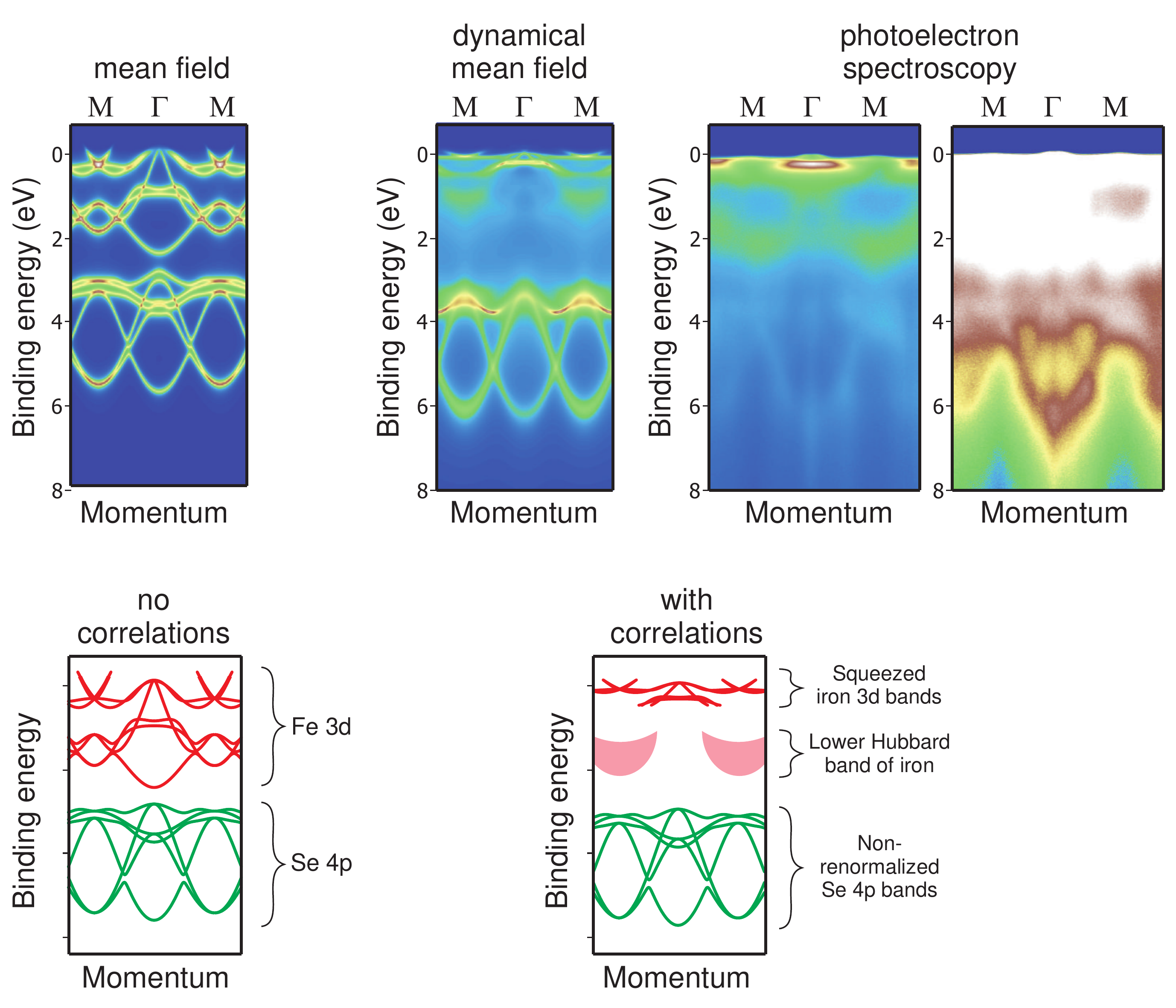}
\vspace{-0.0cm}
\caption{Spectral function of iron and selenium bands in FeSe as seen by LDA, DMFT and ARPES. Top raw: electronic bands obtained in LDA calculations, spectral weight distribution from DMFT and spectra measured by ARPES.  The main deviation of the DMFT and ARPES spectra from LDA is separation of iron 3$d$ band into renormalized bands with clear dispersion at the Fermi level and weakly dispersive lower Hubbard band. At the same time selenium 4$p$ bands are renormalized with respect to LDA neither in DMFT nor in ARPES. Bottom raw: sketch illustrating structure of the electronic spectrum with and without electronic correlations.}
\vspace{-0.0cm}
\label{three}
\end{figure*}

%\begin{figure*}[]
%\includegraphics[width=0.7\textwidth]{Last_figure.pdf}
%\vspace{-0.0cm}
%\caption{}
%\vspace{-0.0cm}
%\label{four}
%\end{figure*}

%\,---\,the puzzle of unconventional high $T_{\rm c}$'s is one of the greatest in the contemporary physics

\noindent Mean field approach to the conduction electrons \cite{MFT} became a classical theory, and is one of the pillars on which the breakthrough of technology in the 20th century resides. At the same time the potential of the electronic systems, obedient to the pure band theory, seems to be largely depleted by now.  As opposed to the case of negligible correlation effects, the Hubbard model \cite{Hubbard} is known to provide a solution in the polar extreme of dominating electronic correlations (see Fig. 1). Interesting properties of many new materials in the scope of the modern condensed matter physics are often found for the regime lying in between normal band metal and correlated insulator \cite{Dagotto, Tokura}. However there is little information as for the actual degree of the electron-electron interaction strength for relevant materials as well as for comprehensive theoretical treatment of the problem.

Electronic correlations were conjectured to be vital for abnormally high critical temperatures in unconventional superconductors \cite{Anderson, Thread, Maier, Haule, Raghu}. In best-known material class, exhibiting highest transition temperatures,\,---\,cuprates\,---\,the superconductivity rises when additional charge carriers are doped into the antiferromagnetic Mott insulator \cite{Fujita, Lee}, immediately invoking the suggestion that correlations are relevant. Although many subclasses of iron-based superconductors exhibit a phase diagram very similar to cuprates \cite{Kamihara, Stewart, Paglione}, there is no evidence for insulating behaviour at any doping level, questioning the degree of correlation strength and their importance in this case. At the same time various experimental techniques have shown that the electronic states at the Fermi level are renormalized three and more times with respect to the predictions of the local density approximation theory (LDA) \cite{Popovich, Sergei_LiFeAs, Cui_NFCA, Orbitalgap, Terashima, Maletz}. Encouragingly similar estimate for band renormalization was obtained in the dynamical mean field theory (DMFT) calculations \cite{Nekrasov, Aichhorn, Valenti, Kotliar, Aichhorn2}. However, it is still not entirely clear to which extent  the underlying physics was captured by DMFT, as renormalization of the low-energy spectrum itself is a rather universal effect, and, for example, also encapsulates the contribution from electronic interactions with low-energy bosonic modes. 

Here we present angle-resolved photoemission spectroscopy (ARPES) measurements of the electronic spectrum of the simplest iron-based superconductor with potentially highest $T_{\rm c}$  \cite{Ge, Feng}, iron selenide. The experimental spectrum of the iron $3d$ band deviates strongly from the LDA, but exhibits a full-scale match with spectral function obtained by DMFT. Remarkably, along with the sharp Fermi-liquid-like bands at the Fermi level, the well-defined dispersive lower Hubbard band (LHB) is present both in calculated and measured spectrum.

Energy-momentum cuts through the photoemission intensity distribution measured in high symmetry directions for binding energies, $\omega$, up to 4\,eV are presented in Fig.~2 along with corresponding plots of the spectral weight distribution obtained in DMFT calculations. Both in ARPES and DMFT the electronic spectrum consists of two parts: (1) renormalized bands with sharp well defined dispersion at the Fermi level and (2) diffuse spectral weight with weak but noticeable dispersion at higher binding energies, with (1) and (2) being separated by a stripe of spectral weight depletion. Such spectral weight distribution is not compatible with single-electron model, while it is common for the solution of the Hubbard model, where (1) is called coherent spectral weight, and (2) is the lower Hubbard band. Details of the photoemission spectrum depend on experimental conditions\,---\,photon energy, light polarization, position in the momentum space, as relative photoemission matrix element for different bands can vary in a wide range. Nevertheless, the most prominent and important features persist and match the counterparts in the calculation, implying that DMFT captures all relevant peculiarities of the spectrum. The position of the LHB in the calculated spectral function is subject to the ill-posed analytic continuation problem. In the supplementary material we show how different variants of analytic continuation can change the position of the LHB such that it agrees very well with the experimentally observed one in terms of location and broadening. It is interesting to note that a model for the spectral function with an empirical self-energy of the simplest shape can reproduce the data satisfactorily well \cite{NaFeAs}, however it fails to reproduce the experimentally observed well-defined LHB separated by a sharp stripe of spectral weight depletion, while these features are inherent to the DMFT calculations.   

%There is a noticeable difference as for position of the LHB observed by ARPES as compared to the predicted by DMFT: LHB in ARPES is centred at about 2\,eV of binding energy, while in the presented DMFT spectra it is centred at about 1.5\,eV. We believe that this difference is not principal, as the position of the LHB in the calculations can be adjusted by variation of the input parameters in the reasonable frames, as shown in Supplementary Materials.

Although the original paper by Hubbard presents Hubbard bands as possessing energy dispersion reminiscent of the unperturbed band \cite{Hubbard}, commonly they are depicted and thought of as non-dispersive spectral weight. The LHB, that we observe here, indeed exhibits large electron scattering rate and is not sharp. At the same time we do observe appreciable dispersion, equally in ARPES and DMFT: the LHB has minimum and is defined best at the M point, it disperses upwards and looses intensity when departing from the M point in any direction. One of the possible interpretations of the observed picture is that LHB emerges out of the states that would be located at the bottom of the iron band in absence of correlations.

%A further analysis of the LHB in the theoretical data reveals an interesting orbital effect: indeed, most of the Fe-$d$ spectral weight above 1\,eV of binding energy stems from the $xy$ and the $x^2-y^2$ orbitals, and the dispersion of the LHB is largely dominated by $x^2-y^2$. This might appear puzzling at first sight, since the latter is\,---\,thanks to its hybridisation pseudo-gap at the Fermi level\,---\,among the least correlated orbitals in this compound. In Supplementary Materials we demonstrate that this is not a contradiction: on the contrary, as we show there, the hybridization gap facilitates the detection of the LHB since it pushes the overall feature to higher binding energies, separating it better from the remaining spectral weight.

We analysed the orbital content of the lower Hubbard band in the calculated spectrum, and found non-zero con-
tributions from all the orbitals. In particular, also the $e_g$-like orbitals ($z^2$ and $x^2-y^2$), contribute, and their contribution is also located at higher binding energies as compared to the most correlated $xy$ orbitals. This might appear puzzling at first sight, since the $e_g$ orbitals are\,---\,thanks to its hybridisation pseudo-gap at the Fermi level\,---\,significantly less correlated than the $t_{2g}$ orbitals. In the supplementary material, we demonstrate that this is not a contradiction: on the contrary, as we show there, the hybridization gap facilitates the detection of the LHB since it pushes the overall feature to lower energies, separating it better from the remaining spectral weight. In contrast, for the $xy$ orbital the Hubbard band is very close to the quasiparticle excitations around the Fermi level and makes it more difficult to separate.

It is instructive to compare spectral functions obtained by LDA, DMFT and ARPES in even wider energy range, encompassing both iron and selenium bands. In Fig.~3 we present spectra in the range up to 8\,eV of binding energy. Interestingly, the bands, predominantly derived from the selenium orbitals, exhibit little deviation from the LDA calculation, which contrasts the described above behavior of the iron-derived bands.  Selenium 4$p$ bands both in ARPES and DMFT are not renormalized with respect to LDA and exhibit moderate electron scattering. Still, one can point out that both in DMFT and in ARPES they are consistently located at a higher binding energy, as compared to the LDA. Very low intensity is explained by drastically different cross-sections of iron $3d$ and selenium $4p$ shells in the photoemission process; in order to compensate this effect and to make selenium-derived bands better visible, we plot ARPES data in a corresponding color scale (last panel in the top row of Fig.~3). Also refer to  the Supplementary Materials for better visualization of the dispersion. 

In conclusion, using state-of-the-art ARPES and DMFT, we have provided evidence for a highly dispersive LHB in the\,---\,probably most intriguing\,---\,representative of iron-based superconductors, FeSe. While the presence of strong Hund's coupling is generally believed to prevent the formation of Hubbard bands, our study reveals a highly dispersive lower Hubbard band, following roughly the non-interacting band dispersion. Our work has implications for other iron-based superconductors, where one may speculate that HBs should be present. It also underlines that the physics of these materials is not only determined by Hund's coupling, but that Coulomb interactions quite generally, and naturally, play a crucial role, too.
 
%Additionally we show that the less-correlated $e_g$ orbitals contribute substantially to this LHB, which is counter-intuitive at first sight but possible due to their large hybridisation gap.
 
Y.S. acknowledges funding from the Wenner-Gren foundation. 
  
%Both DMFT and ARPES indicate qualitatively same deviation of the electronic spectrum from the LDA.

%All unconventional superconductors fall into intermediate level of correlation strength with such characteristic features of the electronic spectrum as presence of both coherent spectral weight in a form of renormalized sharp bands and weakly dispersive diffuse spectral weight, referred to as lower Hubbard band. 

\clearpage

\setcounter{figure}{0}

\renewcommand{\figurename}{Fig. S}

\begin{center} \LARGE
\textbf{Supplementary materials}
\end{center}

\section{Fermi surface map and superconducting transition in ARPES of FeSe}

\noindent In Fig.~S1 one can see ARPES data recorded from freshly cleaved surface of single crystalline sample of FeSe. A Fermi surface (FS) map and a temperature dependence of the integrated energy distribution curve confirming superconducting transition in the studied matter are presented. The Fermi surface consists of hole-like pocket at the Brillouin zone (BZ) center, $\Gamma$ point, and of electron-like pocket at the BZ corner, M point. In the region of low binding energies the electronic bands, supporting the Fermi surface, are very reminiscent of the ones obtained in LDA. The major difference is that experimentally observed bands are renormalized by factor of three and more with respect to the calculation. Temperature-dependent measurements reveal the growth of the well-defined coherence peak, showing the opening of the superconducting gap below transition temperature of about 8\,K. 

% \cite{Maletz}.

\begin{figure}[h]
\includegraphics[width=1\columnwidth]{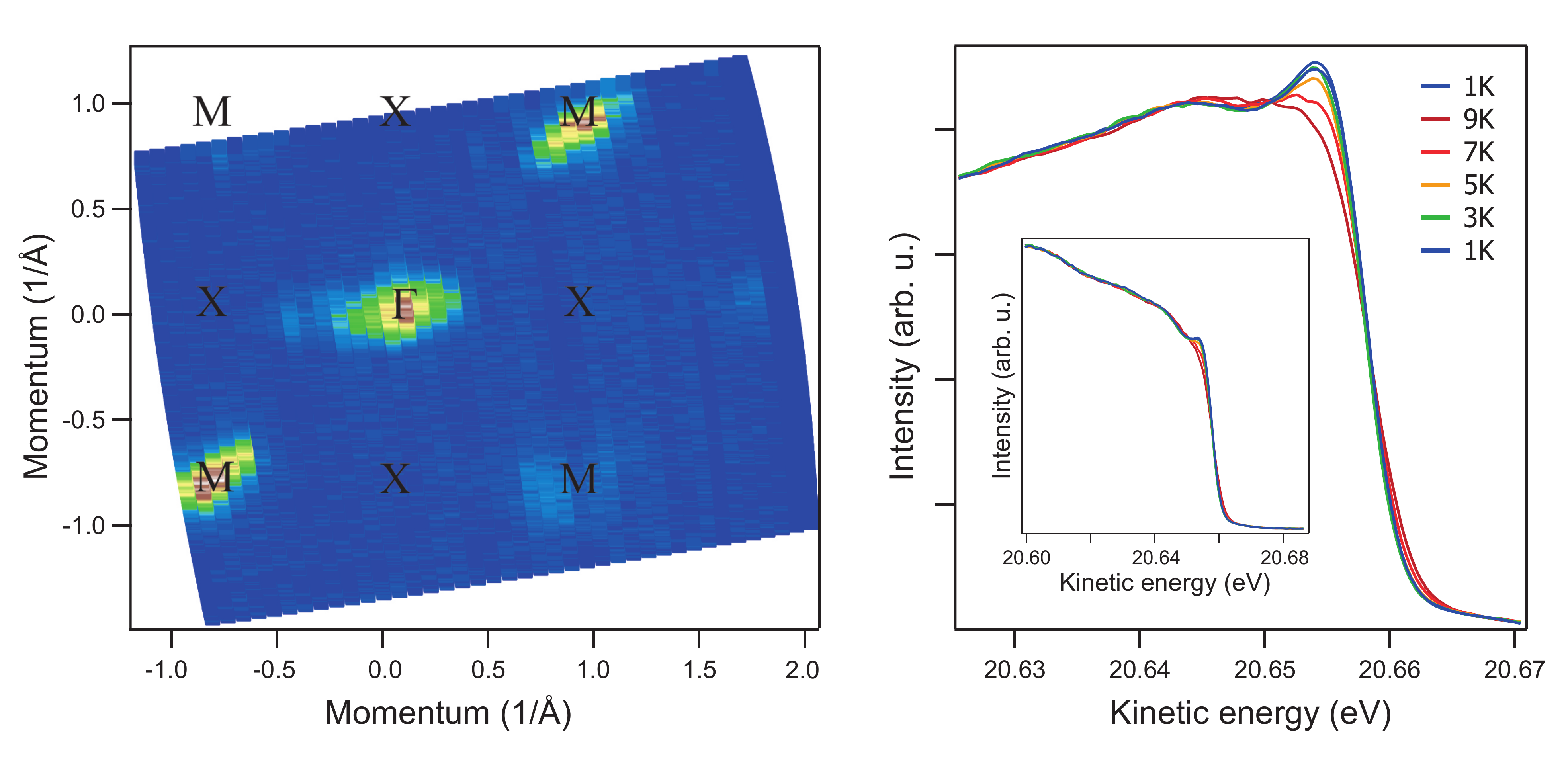}
\vspace{-0.0cm}
\caption{Fermi surface map and observation of the superconducting transition in ARPES spectra of single crystals of FeSe. Left: distribution of photoemission intensity at the Fermi level recorded at photon energy of 80\,eV confirms hole-like Fermi surface sheets around the $\Gamma$ point and electron-like ones at the M point. Right: temperature dependence of the partial density of states referring to the $\Gamma$ pocket reveals growth of the coherence peak below $T_{\rm c}$ of about 8\,K. Spectra in a wider energy range are shown in the inset.}
\vspace{-0.0cm}
\label{one}
\end{figure}

\section{Dispersion of Se 4p bands from ARPES}

\noindent Observation of the clear dispersion of the $4p$ bands of selenium (Fig. S2) in ARPES data ensures generally good agreement with band structure calculation in a wide range of binding energy, from 0 up to 8\,eV. Selenium bands reveal sharp dispersion, relatively small scattering rate, and virtually no renormalization with respect to the LDA calculations. At the same time an interesting observation regarding the difference between LDA and ARPES can be made: the experimental $4p$ bands are shifted towards higher binding energies by about 1\,eV, as compared to the band theory. A similar shift of $4p$ can be observed when DMFT results are matched with LDA, see Fig. 3 of the main text. I the Fig. S2 the color scale is set in a way visualizing the dispersion of the $4p$ bands the best. Photoemission matrix elements vary in a wide range upon changing the experimental conditions, and different parts of the band appear intense and suppressed depending on the photon energy and polarization, and on the electron emission angle. In the panel (a) the results of the LDA calculation are presented, while in the panel (e) only the $4p$ band, shifted to the experimentally observed position, is shown.

\begin{figure}[h]
\includegraphics[width=1\columnwidth]{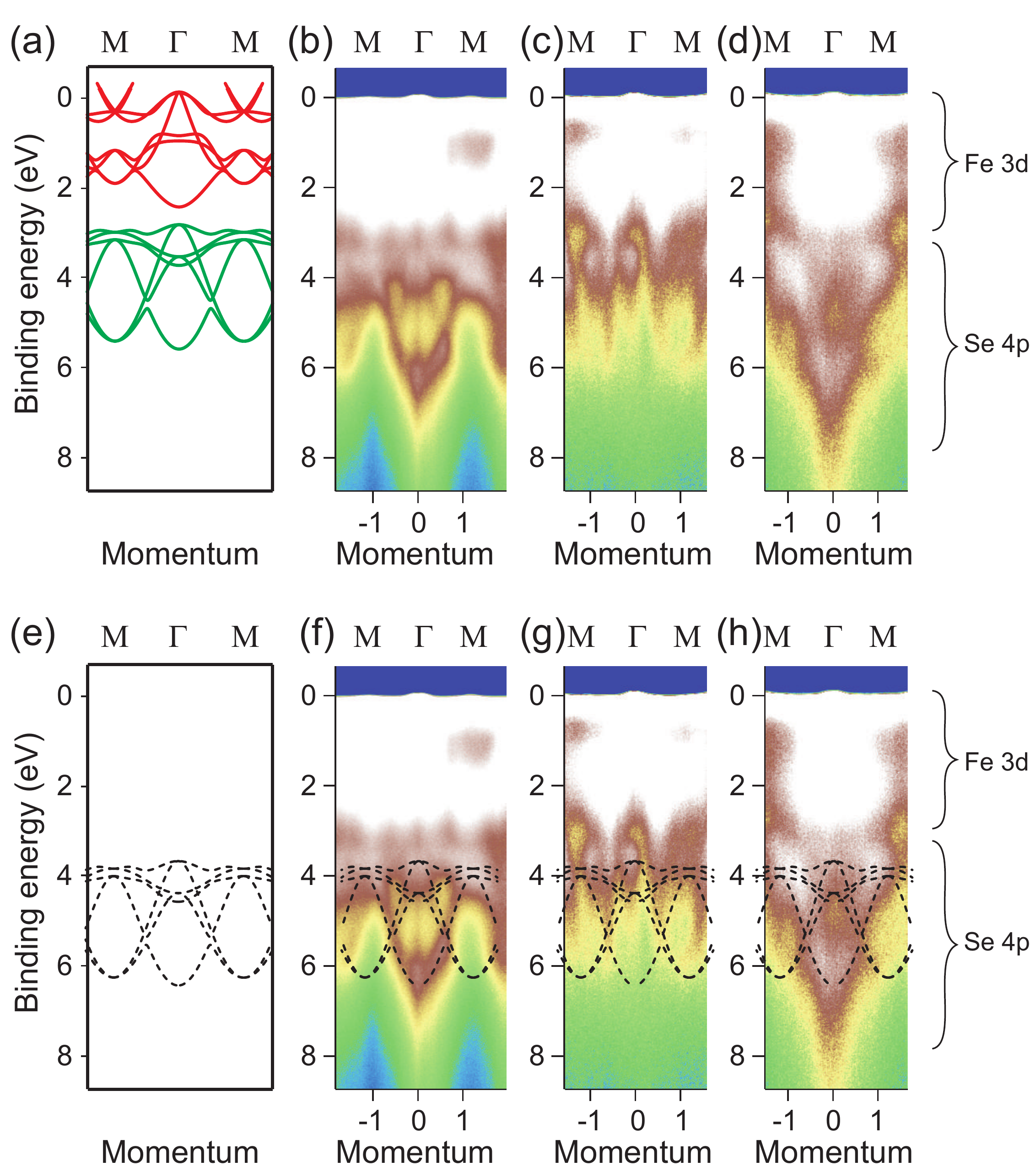}
\vspace{-0.0cm}
\caption{Dispersion for the selenium-derived bands in FeSe. (a) Band dispersion from LDA calculations. (b,c,d) Spectra of FeSe along M$\Gamma$M direction measured with different photon energies and light polarization in order to highlight different parts of the Se 4p band. (e) Position of Se bands corresponding to the one observed experimentally. (f,g,h) same as (b,c,d) with superimposed contours of Se bands.}
\vspace{-0.0cm}
\label{three}
\end{figure}

%\section{Position of Lower Hubbard Band in DMFT spectra}
%\begin{figure*}[]
%\includegraphics[width=0.4\textwidth]{New_old_DMFT.jpg}
%\vspace{-0.0cm}
%\caption{}
%\vspace{-0.0cm}
%\label{U}
%\end{figure*}

\begin{figure*}[]
\includegraphics[width=0.95\textwidth]{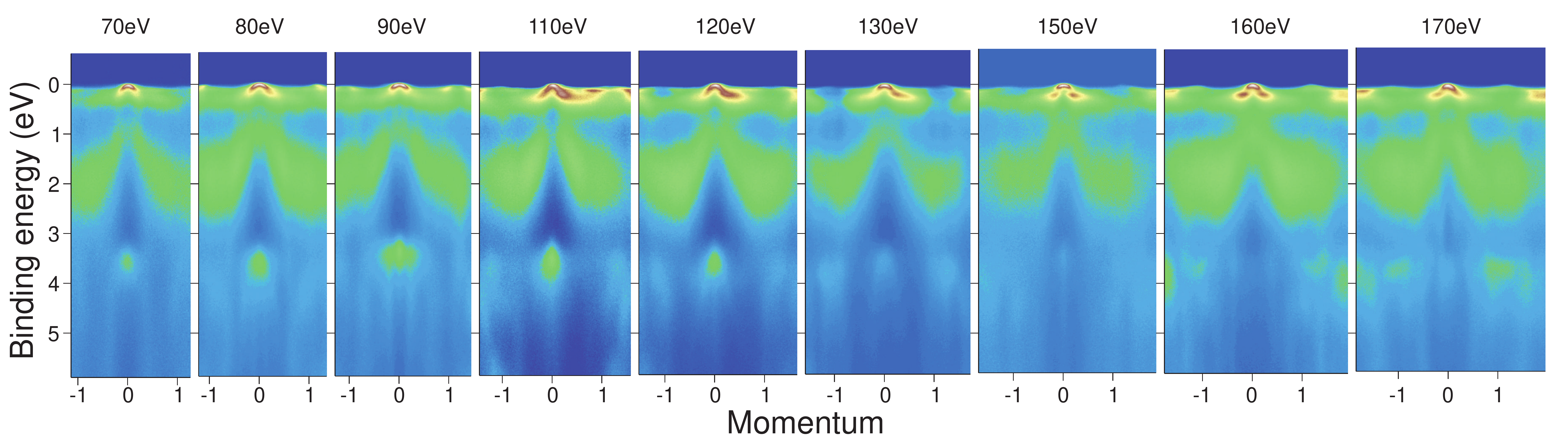}
\vspace{-0.0cm}
\caption{An energy-momentum cut passing through the $\Gamma$ point, recorded at various photon energies and vertical light polarization. The photon energies are indicated above the corresponding intensity plots. One can see that the distribution of photoemission intensity in the region of binding energies 0.5..3\,eV, where the LHB resides, is rather universal, implying that although the matrix elements play important role, one can extract the parameters of the LHB from experiment with good precision.}
\vspace{-0.0cm}
\label{hv}
\end{figure*}

\section{Independence of Lower Hubbard Band position in ARPES as a function of photon energy}

\noindent Photoemission matrix elements can have a strong impact on the measured signal, especially when the spectral function consists of rather broad peaks. In order to reveal the true underlying distribution of the spectral weight, we have performed measurements at different conditions. A series of spectral images, recorded along the direction passing through the Brillouin zone center, the $\Gamma$ point, at different photon energies, are presented in the Fig. S3. The generic features of the spectral weight distribution, such as positions of the coherent spectral weight, LHB, a stripe of the spectral weight depletion between them remain unchanged.

%\newpage

\section{Lower Hubbard Band position in DMFT}

\noindent In the main text of this paper we included DMFT data that is produced in
accordance with Ref.~\onlinecite{Aichhorn2S}, in order to be consistent with published
data. We use a continuous time quantum Monte Carlo technique \cite{SethS} for
the calculations, which produces Greens functions and self energies on
the Matsubara frequency axis. In order to get data on the real-frequency
axis that can be compared to the ARPES experiments, one faces the ill-posed
analytic continuation (AC) problem. In addition, the AC is not defined
to be used directly for self energy. For the data in the main text-and
also in Ref.~\onlinecite{Aichhorn2S}\,---\,we used the following procedure. One defines an
artificial Greens function, ${\tilde G}(i\omega) = \frac{1}{i\omega-\Sigma({i\omega})+\Sigma^{\text{DC}}}$, where
$\Sigma^{\text{DC}}$ is the double counting correction. This ${\tilde G}(i\omega)$ is continued to the real frequency axis, which after inversion of above equation leads to the real frequency
$\Sigma({\omega})$. Since $\Sigma(i\omega)$ is damped by the $1/i\omega$
tail of the Greens function, high energy features of
$\Sigma(\omega)$ are particularly smooth.

\begin{figure}[b]
\includegraphics[width=1\columnwidth]{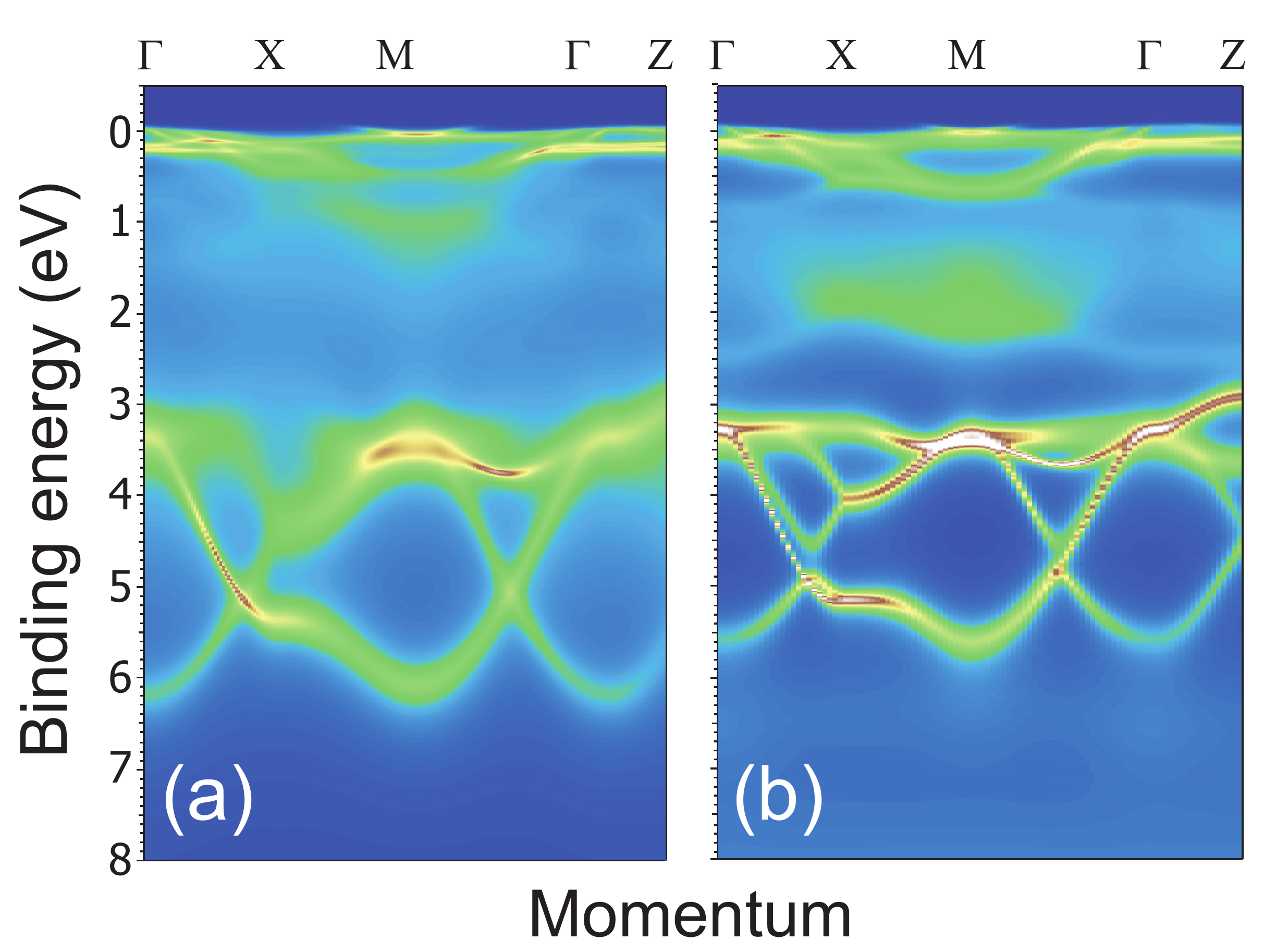}
\vspace{-0.0cm}
\caption{Calculated spectral function using two different variants of the
analytic continuation method for the self energy. Left (a) using an
artifical Greens function, right (b) using a rescaling of the self energy.}
\vspace{-0.0cm}
%\label{newold}
\end{figure}

Here, we analyse the influence of this smoothening by applying another
scheme to do the analytic continuation of the self energy. We modify
$\Sigma(i\omega)$ directly, such that it behaves as a Greens function.
In other words, we define the artificial function
${\tilde\Sigma}(i\omega)$ by substracting the real part of
$\Sigma(i\omega\to\infty)$, and rescaling it such that ${\tilde
\Sigma}(i\omega)$ is normalised to one. Then, one can apply AC directly
to ${\tilde\Sigma}(i\omega)$.

In Fig. S4 we compare the two method, using the artificial Greens
function ${\tilde G}(i\omega)$, left (a), and the rescaled self energy,
right (b). It is immediately obvious that the overall features-low
energy renormalization, spectral weight suppression, lower Hubbard
bands-are similar in both methods. The position and sharpness of the
features, however, differ between the methods, in particular at high
energies. This is not surprising, given the ill-posedness of the AC.
Using the rescaled self energy puts the position of the Hubbard bands
even closer to what is measured in ARPES. Nevertheless, in order to be
consistent with published data, we discuss both variants of doing the
AC, and do not just use one of them.

\section{Influence of the one-particle band structure on the 
dispersion of Hubbard bands: spectral properties of correlated orbitals with
hybridisation gap}

\noindent In this section, we show why the existence of a single-particle pseudogap
in the non-interacting density of states in a multi-orbital system
helps the formation of Hubbard bands in those orbitals. The mechanism
is simple: the separation of the Hubbard bands, which in the atomic
limit is given by the Hubbard $U$ (for simplicity we consider a model
with $J$=0), is substantially enhanced by the hybridization, see
Fig.~S5, and equations below.

\begin{figure}[htb]%
\centering
\includegraphics[width=1\columnwidth]%width=0.1\columnwidth]%,%angle=270,clip=]
{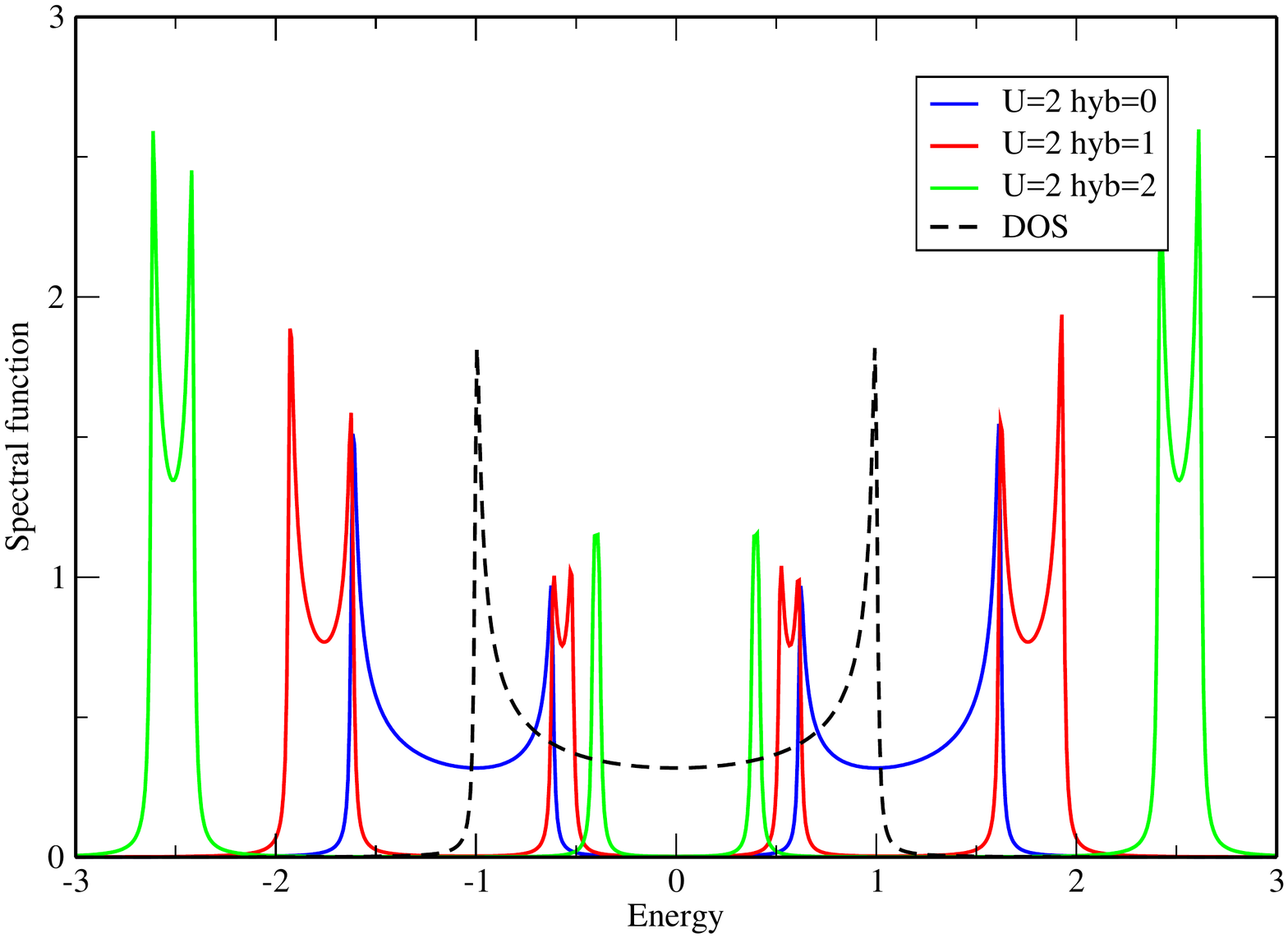}
\caption{Spectral function of a two-orbital model with non-interacting
DOS given by the dashed black lines, calculated within the Hubbard 1
approximation.}
\label{fig1}%
\end{figure}

Let's consider a Hubbard model with two orbitals
with Hamiltonian given by $H_{11} = \epsilon_k = - H_{22}$, $H_{12}=v$.
For the application below, we will take $\epsilon_k$ in the form of
a cosine band, but this is not essential, as long as the two bands
$\epsilon_k$ and $-\epsilon_k$ intersect. The effect of $v$ is then
obviously to open a hybridisation gap.

The eigenvalues are
\begin{eqnarray}
E^{(0)}_{\pm} = \pm \sqrt{\epsilon_k^2 + v^2}
\end{eqnarray}

Now, we add Hubbard interactions {\bf on the original orbitals}.
We use the Hubbard 1 approximation, that is, we assume the self-energy
not to be modified by the hopping:
\begin{eqnarray}
\Sigma(\omega+ i \eta) = \frac{U^2}{ 4 (\omega+i \eta)^2}
\end{eqnarray}

This splits each of the two bands, and the spectral function consists
of poles at energies
\begin{eqnarray}
E_{\pm} &=& \pm \frac{1}{2} E^{(0)}_{\pm} 
\pm \frac{1}{2} 
\sqrt{(E^{(0)}_{\pm})^2 + U^2}
\\
&=&
\pm \frac{1}{2} 
\sqrt{\epsilon_k^2 + v^2}
\pm \frac{1}{2} 
\sqrt{\epsilon_k^2 + v^2 +U^2}
\end{eqnarray}

The two outermost features (Hubbard bands) are separated by
\begin{eqnarray}
\sqrt{\epsilon_k^2 + v^2}
+
\sqrt{\epsilon_k^2 + v^2 +U^2}
\end{eqnarray}
which exceeds both, the non-interacting bonding-antibonding
splitting of $2 \sqrt{\epsilon_k^2 + v^2}$ and the Hubbard
splitting in the absence of hopping/hybridisation $U$.
The effect of the hybridisationg gap is thus to push the Hubbard
features away from the Fermi level, making them easier to detect.
The dispersion of the lower Hubbard bands is then essentially given
by the dispersion of the non-interacting band, shifted to higher
binding energies.

This explains why Hubbard bands in iron-based superconductor
materials are most prone to appear in those bands that are
``pseudo-gapped'' (at the single particle level), resolving the
apparent contradiction of the Hubbard band having substantial
weight on the least correlated orbitals.

\end{document}